\RequirePackage{fix-cm}
\documentclass[twocolumn]{svjour3}       
\smartqed  
\usepackage{graphicx}
\usepackage{adjustbox}
\usepackage{graphicx}
\usepackage{array}
\newcolumntype{N}{>{\centering\arraybackslash}m{.5in}}
\newcolumntype{M}[1]{>{\centering\arraybackslash}m{#1}}
\usepackage{multirow, makecell}

\begin{document}

\title{Contact forces distribution for a granular material from a Monte Carlo study on a single grain
}


\small{\author{Manuel A. C\'ardenas-Barrantes \textit{$^{1}$}         \and
        Jose Daniel Mu\~{n}oz \textit{$^{1}$} 		\and
	William F. Oquendo \textit{$^{2}$}}		 
}


\institute{Manuel A. C\'ardenas Barrantes\at
              \email{macardenasb@unal.edu.co} \\ \\
		\textit{$^1$} Simulation of Physical Systems Group, Department of Physics,
                Universidad Nacional de Colombia, Carrera 30 No. 45-03, Ed. 404, Of. 348, 	Bogota D.C., Colombia.\\
           \and
		\textit{$^2$} Department of Mathematics, Physics, and Statistics, Faculty of Engineering, Universidad de la Sabana, Km 7 Autopista Norte de Bogota, Ch\'ia, Colombia.\\
}

\date{Received: date / Accepted: date}

\maketitle

\begin{abstract}
The force network ensemble is one of the most promising statistical descriptions of granular media, with an entropy accounting for all force configurations at mechanical equilibrium consistent with some external stress. It is possible to define a temperature-like parameter, the {\it angoricity} $\alpha^{-1}$, which under isotropic compression is a scalar variable. This ensemble is frequently studied on  whole packings of grains; however, previous works have shown that spatial correlations can be neglected in many cases, opening the door to studies on a single grain. Our work develops a Monte Carlo method to sample the force ensemble on a single grain at constant angoricity on two and three-dimensional mono-disperse granular systems, both with or without static friction. The results show that, despite the steric exclusions and the constrictions of Coulomb's limit and repulsive normal forces, the pressure per grain always show a gamma distribution with scale parameter $\nu=\alpha^{-1}$ and shape parameter $k$ close to $k'$, the number of degrees of freedom in the system. Moreover, the average pressure per grain fulfills an equipartition theorem $\langle  p  \rangle  = k' {\alpha}^{-1} $ in all cases (in close parallelism with the one for an ideal gas). These results suggest the existence of $k'$ independent random variables (i.e. elementary forces) with identical exponential distributions as the basic elements for describing the force network ensemble at low angoricities under isotropic compression, in analogy with the volume ensemble of granular materials.
\end{abstract}

\section{Introduction}

Granular media, like sand, rice, coffee grains, soils and powders, are relevant in everyday's life and of paramount importance on industrial applications. However, there is still no general theoretical description for them. Most of our knowledge is ciphered in empirical constitutive equations relating the stresses, the strains and the energies inside the material \cite{Jacques2000} through many parameters whose microscopic origins are not always clear. But still there is a beacon of hope in statistical mechanics. Granular media are systems of many particles with a relevant interest in macroscopic quantities like volume fractions, strains and external stresses; so, they would be perfect candidates for a statistical mechanics analysis \cite{Henkes2009,cowin1979}. The task, nevertheless, is not easy, because granular media are dissipative systems. One major statistical mechanics approach is the {\it force network ensemble}  \cite{Hecke2004,Vlugt2011,Ellenbroek2004}, which considers all possible configurations on a fixed contact network among grains that fulfill the requirements of mechanical equilibrium in agreement with some imposed external stress. Many configurations are possible, because the number of variables (contact forces) usually is greater than the constraints (equations of mechanical equilibrium per grain and values of external stress), in what is called {\it hyperstaticity}. A force entropy is defined by counting all possible force networks compatible with the external stress, and a temperature-like quantity, the {\it angoricity} is defined to relate that entropy with that external stress \cite{Edwards1998}. For the case of isotropic compression, this external stress is characterized by a single parameter, the pressure $p$, and the angoricity is a scalar variable, $\alpha^{-1}$.

There are many theoretical, computational and experimental works on the force network ensemble. For instance, F. Radjai and coworkers \cite{Radjai1990} found by contact dynamics simulations that  forces below the mean distribute like a power law, and forces above, like an exponential decay. Also, Metzger and Donahue \cite{Metzger2005} suggested that forces follow a Bose-Einstein distribution. Later on,  Thige, Snoejder, Vlugh and coworkers derive from dimensional considerations an equipartition relation for frictionless grains, which describes the expected value of the pressure as a function of the excess correlation number and the angoricity \cite{Vlugt2011}. By using a clever Monte Carlo algorithm (the {\it wheel moves}) to sample a 2D triangular array of monodisperse disks under isotropic compression \cite{Schaeffer2005}, they found that  the correlation length in such a system is of the order of a single diameter \cite{Ellenbroek2004}, and that correlations among grains can be neglected in many cases \cite{Vlugt2011}. This exciting result suggests that the main behavior of the entire system could be understood from the behavior of the forces on a single grain. Indeed, the granocentric model proposed by Maxime Clusel and co-workers \cite{Clusel2009,Corwin2010,Puckett2011} manages to capture the essential properties of the dense granular material, such as the global density, from the statistics of the available space and the ratio of contacts to neighbors around a single grain, obtained from experimental measurements.

The present work develops a Monte Carlo procedure to sample the force network ensemble on a single grain, which is assumed to be part of a mono disperse granular system, both in 2D and 3D, with either frictionless or frictional interactions, and uses this method to investigate the ensemble. The method does not fix a pressure on the grain, but an angoricity (canonical ensemble), and only accepts mechanically stable configurations obeying steric exclusions and using non-cohesive normal forces and frictional forces (if any) below the Coulomb threshold. In all cases studied, the pressure follows a gamma distribution and the angoricity and pressure per grain fulfill an equipartition relation that, when extrapolated, reproduces the functional relation found by Thige and Vlugh for the whole system \cite{Vlugt2011}. This would validate the hypothesis that, under certain conditions, the statistics of forces on each grain can be considered as independent and shows that steric exclusions and force constraints do not have a measurable effect on these results.

\section{The force network ensemble}

Following Thige and Vlugh \cite{Vlugt2011}, let us consider a granular sample of $N$ grains in a two-dimensional box, in quasi-mechanical equilibrium with some external stress. The statements of mechanical equilibrium per grain and consistency with an external stress $\hat{\sigma}$ are given by
\begin{eqnarray} \label{eqfor}
 \forall i, \sum_{j}{\vec{f_{ij}}=\vec{0}} \nonumber  \quad, \\
 \hat{\sigma}=\frac{1}{2V}\sum_{ij}{\vec{f_{ij}}\otimes\vec{r_{ij}}} \quad, 
\end{eqnarray}
where $\vec {f_{ij}}$ is the contact force acting on grain i by grain j,  $\vec {r_{ij}}$ is the vector from the center of grain $i$ to the center of grain $j$ and $V$ is the volume (two-dimensional volume) of the packing. Eqs. (\ref{eqfor}) can be expressed in matrix form as
\begin{eqnarray}  \label{eqmat}
 \textbf{A}f=0 \nonumber \quad ,  \\
 \textbf{B}f=b \quad.
\end{eqnarray}
Both arrays $A$ and $B$ are defined by the packing's geometry. The vector $b$ gathers the independent components of the stress tensor, $b=(\sigma_{xx},\sigma_{yy}, \sigma_{xy})$, and vector $f$ includes all $N_c$ components of the interacting forces among grains (that is, the number of variables).
Besides, one says that a contact network is isostatic if there is only one force network satisfying Eq. (\ref{eqmat}), and hyper-static otherwise.
Under isotropic pressure, $b(p)=(-p,-p,0)$, the density of states for a fixed value of pressure $p_{0}$ can be obtained as
\begin{equation} \label{eqden} 
 \Omega(p)=\int\prod_{if}{\vec{df_{ij}}\delta(\textbf{A}f)\delta(\textbf{B}f-b(p))\Theta(\vec{f_{ij}})}\quad,
\end{equation}
where the integral runs over all contact forces. The three factors $\delta(\textbf{A}f)$, $\delta(\textbf{B}f-b(p))$ and $\Theta(\vec{f_{ij}})$ assure for the mechanical constraints. The Heaviside step function $\Theta(\vec{f_{ij}})$ assures that the contact forces are repulsive ones.

The hyperstaticity can be characterized by the mean excess coordination number, $\Delta z:=\langle z \rangle-\langle z\rangle_{iso}$. For instance, let us consider frictionless systems in 2D. In this case, the total number of equations is $2N + 3$. Thus, the system is hyper-static when $N_c>2N + 3$,  where $\langle z\rangle_{iso}=4+\frac{6}{N}$ ($\simeq 4$ for large systems) is the coordination number of the isostatic system (assuming no periodic boundary conditions). Since each contact has a single variable shared by two grains, the number of excess variables is just $N_W=\frac{1}{2}N\Delta z$.

When $\Delta z>0$, any force network $f$ satisfying Eq. (\ref{eqfor}) can be written as \cite{Schaeffer2005}
\begin{equation} \label{eqfor2} 
 f=f_0+\sum_{k=1}^{N_W}w_k\delta f_k \quad,
\end{equation}
where $f_0$ is a particular solution to Eq. (\ref{eqmat}), $\delta f_k$ are base vectors spanning the null space of solutions and $w_k $ are the components of $f$ in such a base. For 2D frictionless systems, for instance, the dimension of the space of all possible force networks is $N_W=\frac{1}{2}N\Delta z$. 

The maximum entropy principle can also be used to build up a canonical ensemble, with entropy
\begin{equation} \label{eqent} 
 S[B]=-\int{dfG(f)[ln(B(f))B(f)]}\quad ,
\end{equation}
where $G(f)$ takes values of 1 or 0 if the force network fulfills or not  the mechanical constraints. Under isotropic compression, the external stress is characterized by a single parameter, the external pressure $p$. A force with magnitude $f$ appears in this canonical ensemble with probability 
$B(f)=\frac{e^{-\alpha p(f)}}{Z}$, with $Z(\alpha)=\int{dfG(f)e^{-\alpha p(f)}}=\int{dp \Omega(p)e^{-\alpha p}}$ the partition function and $\alpha$, a scalar Lagrange multiplier, whose  inverse is called {\it angoricity} \cite{Hecke2004}. The average pressure can be computed as usual, 
\begin{equation} \label{eqpre2} 
  	\langle p\rangle=-\frac{\partial}{\partial \alpha} \ln Z \quad .
\end{equation}
Because the space of force networks is a convex polytope in $N_w$ dimensions, with $p$  the linear dimension of the polytope, one can assume \cite{Vlugt2011} $\Omega (p)\propto p^{N_W}$; therefore, by Eq. (\ref{eqpre2}),
\begin{equation} \label{eqsta} 
  	\langle p\rangle\alpha\simeq N_W\quad,
\end{equation} 
in the thermodynamic limit.

From a microscopic point of view, one can define a local pressure on a single grain $p_i$ as the sum of all normal forces acting on it. The density of states for this pressure is
\begin{equation} \label{eqdenpi} 
 \Omega(p_i)=\int{\vec{df_{ij}}\delta(\textbf{A}f_j)\delta(\textbf{B}f_j-b(p_i))\Theta(\vec{f_{ij}})},
\end{equation} 
where $\delta(\textbf{A}f_j)$, $\delta(\textbf{B}f-b(p_i))$ and $\Theta(\vec{f_{ij}})$ assure for mechanical equilibrium, external stress and normal repulsive forces, as before.
For a single grain with frictionless contacts in $d$ dimensions, for instance, $f$ is a vector of $z$ components (that is, the number of variables is $z$) and the number of mechanical constraints is $d+1$ (one per dimension, to assure equilibrium, plus one to fix $p_i$). Thus, the configuration space at fixed pressure $p_i$ has $N_i=z_i -d -1$ dimensions \cite{Vlugt2011}, and one could assume 
\begin{equation} \label{eqdenpi2} 
 \Omega(p_i)\propto p_i^{z_i-d-1}\quad,
\end{equation} 
which has  been verified by molecular dynamics simulations \cite{Hecke2004}.
Then, from the equipartition function for a single grain, 
\begin{equation} \label{eqentrpi2} 
 Z_i(\alpha)=\int_0^\infty dp_i \Omega(p_i) e^{-\alpha p_i}\quad,
\end{equation} 
one obtains Eq. (\ref{eqpre2}),
\begin{equation} \label{eqsta} 
  	\langle p_i\rangle\alpha= N_i+1=z_i-d \quad.
\end{equation} 
Similar results follow for frictional contacts or 3D ensembles, but with other values of $N_i$. These are the relations we would like to verify by using Monte Carlo.

\section{2D-Monte Carlo sampling on a single grain}

\subsection{2D Frictionless grains}

Consider a single 2d-grain with z contacts, which is assumed to be part of a two-dimensional mono-disperse set of frictionless disks. The Monte Carlo sampling starts by setting the angular positions $\theta_i$  ($i=0,1,\cdots,z-1$) for the z contacts. The main restriction here is steric exclusion, which also limits the coordination number z to a maximum of six. In addition, because we are just interested in hyperstatic systems, the minimum coordination number z is three, otherwise the system would be completely defined by the constraint of mechanical equilibrium. We consider two different possibilities for the contacts: either at fixed positions, equally spaced across the circle (called {\it fixed}) or randomly chosen from all possible configurations obeying steric exclusion (that is, with angles larger than $\pi/3$ between consecutive contacts) and using $100$ contact configurations per run. 
Next, we assign an initial set $\{f_{i}\}$ of non-cohesive forces $(f_i > 0)$ for the contacts satisfying the constraints Eq. (\ref{eqfor}) of mechanical equilibrium. This is done by choosing $z-2$ contacts at random and choosing $f_i = 1$ for them. The normal forces for the other two contacts are obtained from the mechanical equilibrium equations Eq. (\ref{eqfor}),

\begin{eqnarray}  \label{eqequ2}
 \sum_{i=1}^zf_icos(\theta_i)=0 \nonumber \quad ,  \\
 \sum_{i=1}^zf_isin(\theta_i)=0 \quad.
\end{eqnarray}
  
The pressure on the grain is computed as $p=\sum_i f_i$.

Once the initial configuration is set, we start a Metropolis sampling scheme at constant angoricity, as follows:
A new force configuration is generated by choosing $z-2$ contacts at random and changing the value of its normal force in a small amount $\Delta f_i$, randomly generated on the interval $[-a, a]$ ($a=0.1$ for our simulations). The other two normal forces are modified to assure that mechanical equilibrium Eq. (\ref{eqequ2}) is fulfilled. Then, the restrictions of non-cohesive forces are checked: If any new force is negative, the force configuration is rejected and a new one is randomly generated; otherwise, the new value of the pressure is computed. At this point, we introduce the Metropolis acceptance rate: If the pressure change $\Delta = p_{new}-p_{old} \le 0$, the move to the new configuration is accepted; otherwise, it is accepted only if a random number $r\in [0, 1]$ is less or equal than $e^{(-\alpha\Delta p)}$. This process is repeated as many times as necessary to reach equilibrium, i.e. when the pressure starts to fluctuate around a mean value. Typical equilibrium times are around $10^6$ time steps for $\alpha^{-1}=1/6$, and lower for higher angoricities. Sampling starts after two equilibrium times. Once equilibrium is reached, the time correlation function for the pressure was computed and the correlation time, $t_{corr}$, measured. Typical correlation times were around $10^4$ for $\alpha^{-1}=1/6$ and smaller for higher angoricities. Samples are taken every $2 t_{corr}$.

\begin{figure}[h]
\centering
  \includegraphics[width=8cm]{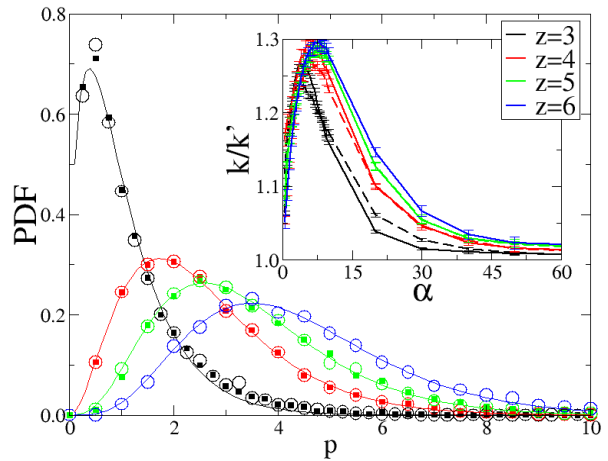}
  \caption{Probability distribution of the pressure on a single grain in a monodisperse frictionless 2D-system with  angoricity $\alpha^{-1}=0.025$ for both contacts at random (filled squares) or fixed (empty circles) positions. Solid lines are Gamma functions with shape parameters $k$ form Eq. (\ref{eqksigma}). (inset) Excess in shape parameter, $k-k'$ as a function of the inverse angoricity $\alpha$, with $k'=z-2$ the number of degrees of freedom in the system, for both random (solid lines) and fixed (dashed lines) contact positions.}
  \label{fgr:PDF1}
\end{figure}

\begin{figure}[h]
\centering
  \includegraphics[width=8cm]{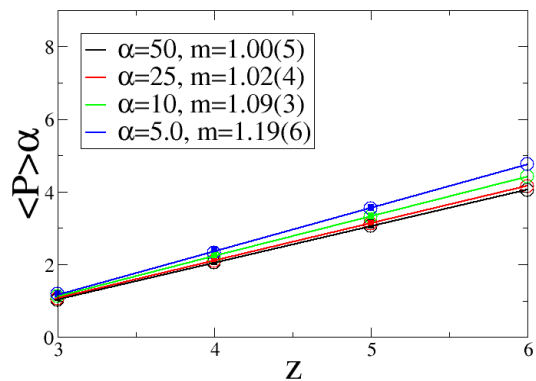}
  \caption{Average pressure $\left< p \right>$ times the inverse angoricity ($\alpha$) as a function of the coordination number $z$  for a grain in a monodisperse frictionless granular 2D-system for both fixed (empty circles) and random (filled squares) contact positions. Here, $m$ is the fitted slope.}
  \label{fgr:pa_z1}
\end{figure}

\textbf{Fig. \ref{fgr:PDF1}} shows the pressure distribution for different values of angoricity and contact numbers (with both fixed and random contact positions). All curves  are well fitted by Gamma distributions,
\begin{equation} \label{eqdis}
  \Phi(p)=\frac{1}{\Gamma(k)v^k}p^{k-1}e^{-\frac{p}{\nu}} \quad,
\end{equation}
where $k$, known as the {\it shape parameter}, and $\nu$, known as the {\it scale parameter}, are estimated from the average pressure $\left< p \right>$ and the variance $\sigma_p^2$ in the histogram as
\begin{equation}\label{eqksigma}
	k=\frac{ \langle p\rangle^2}{\sigma_p^2}\quad , \quad  \nu=\frac{\sigma_p^2}{\langle p\rangle}\quad.
\end{equation}
By plotting $k$ against $\alpha$ (\textbf{Fig. \ref{fgr:PDF1}}, inset) we found  that the parameter $k$ reaches a maximum for low $\alpha$ and starts to decay. For low angoricities (large $\alpha$ values) $k\to k'=z-2$, which is the number of degrees of freedom in this one-grain system. From this result and Eq. (\ref{eqdis}), we conclude that the  the pressure on a single frictionless grain distributes like
\begin{equation} \label{eqdisp1}
  \Phi(p)\propto p^{k'-1} =p^{z-2-1}\quad .
\end{equation}
This is the same relation Eq. (\ref{eqdenpi2}) found by Tighe B P, Vlugt T. in \cite{Vlugt2011} for a bidimensional force packing.

Plotting $\langle p\rangle \alpha$  against $z$ (\textbf{Fig. \ref{fgr:pa_z1}}) shows a linear relationship between these two quantities for all values of angoricity,
\begin{equation} \label{eqest2}
   	\langle p\rangle=(z-2)\alpha^{-1}\quad,
\end{equation} 
and the relation is even better for low angoricities. Equation (\ref{eqest2}) is a clear representation of an equipartition theorem,  in perfect parallel with the equipartition relation for an ideal gas, where the average pressure plays the role of the energy, and the angoricity, the one of temperature. This result can be extrapolated to the whole packing if one takes the number of degrees of freedom as the dimensionality of the space of hyperstatic solutions, $\frac{1}{2}N (z-2)$, recovering the equipartition relation Eq. (\ref{eqsta}) proposed by Thighe and Vlugh.

\subsection{2D Grains with static friction} \label{GSF_2D}

\begin{figure}[h]
\centering
  \includegraphics[width=8cm]{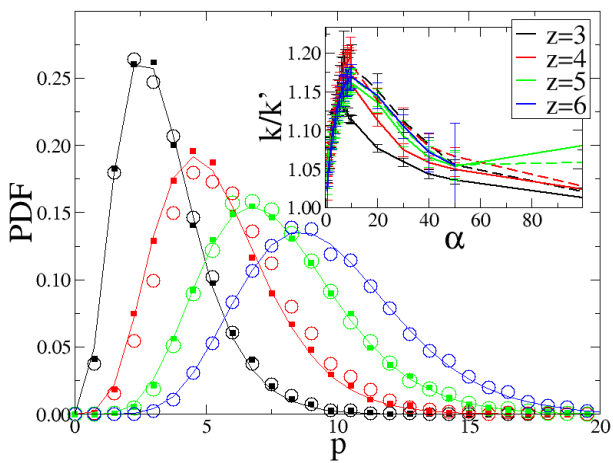}
  \caption{Probability distribution of the pressure on a single grain in a monodisperse  2D-system with static friction coefficient $\mu=0.5$ and angoricity $\alpha^{-1}=0.025$. (inset) Excess in shape parameter, $k-k'$ as a function of the inverse angoricity $\alpha$, with $k'=2z-3$. Symbols and line styles as in \textbf{Fig. \ref{fgr:PDF1}}.}
  \label{fgr:PDF2}
\end{figure}

To extend the previous analysis to a more realistic situation, grains with static friction, only few changes must be introduced. First, there are now two variables per contact: the normal $f^{(n)}_i$ and tangential $f^{(t)}_i$ components of the contact force. Second, the conditions for mechanical equilibrium are now three: two for the forces plus one for the torques,
\begin{equation}
  \sum_{i=1}^z f^{(n)}_i cos(\theta_i)-f^{(t)}_i sin(\theta_i)=0, \nonumber
\end{equation}
\begin{equation} 
  \sum_{i=1}^z f^{(n)}_i sin(\theta_i)+f^{(t)}_i cos(\theta_i)=0, \nonumber
\end{equation} 
\begin{equation} \label{eqequ3}
  \sum f^{(t)}_i=0\quad.
\end{equation}
Third, there is an extra restriction: The absolute value of the tangential force at each contact cannot surpass the Coulomb limit, $f^{(t)}\le \mu f^{(n)}$. The torque condition in the initial configuration is assured by starting with all $f^{(t)}_i=0$ and solving for two normal random forces, as before. The equilibrium conditions Eq. (\ref{eqequ3}) are assured at each Monte Carlo step by solving for three random variables, instead of two. The small variations in the tangential forces are randomly generated from the interval $[-0.01,0.01]$ to avoid an extreme number of rejections, that is ten times smaller than the one for the normal forces. The Coulomb restriction is included when checking for non-cohesive normal forces. Anything else is unchanged.
	
We see again a gamma distribution for the probability in all cases (\textbf{Fig. \ref{fgr:PDF2}}). The parameter k shows the same behavior as before, but reaching a different value $k'=2z-3$ for low angoricities, that is the new number of degrees of freedom in the system. Thus, the probability function for a grain is now proportional to 
\begin{equation} \label{eqdisp2}
  \Phi(p)\propto p^{k'-1} =p^{2z-3-1}\quad ,
\end{equation}
which extends the relationship found by Tighe and Vlugt \cite{Vlugt2011} Eq. (\ref{eqdenpi2}) to the frictional case.

\begin{figure}[h]
\centering
  \includegraphics[width=8cm]{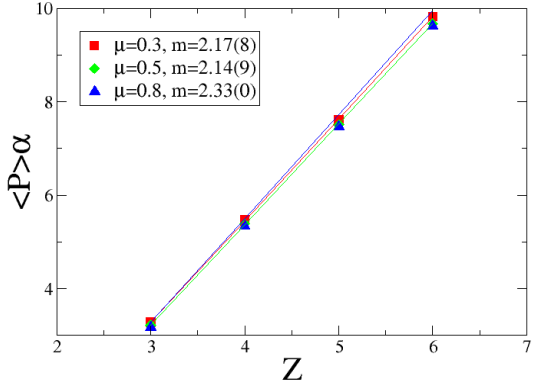}
  \caption{Average pressure $\left< p \right>$ times the inverse angoricity $\alpha$ as a function of the coordination number $z$ for a grain in a mono disperse 2D granular system with friction and angoricity equals to 0.02. Contact positions are chosen randomly for $z=3,4,5$ and fixed  for $z=6$. The figure shows results for static friction coefficients $\mu=0.3$ (squares),  $\mu=0.5$ (diamonds) and  $\mu=0.8$ (triangles).} 
  \label{fgr:pa_zf}
\end{figure}

Plotting the average pressure $\left< p \right>$ times the inverse angoricity $\alpha$ against the coordination number (\textbf{Fig. \ref{fgr:pa_zf}}) evidences again an equipartition equation relating these two quantities, 
\begin{equation} \label{eqsta3}
   	\langle p\rangle=(2z-3)\alpha^{-1} \quad,
\end{equation}
where just the number of degrees of freedom has changed. Furthermore, Eq. (\ref{eqsta3}) shows to be the same for all simulated friction coefficients.

\section{3D-Monte Carlo sampling on a single grain}

The Monte Carlo procedure on a single three-dimensional grain runs very similar to the two-dimensional case. The first difference is that the $z$ contacts are chosen on the sphere, either in regular configurations or at random positions. In this case, once the first contact is set, the second one is accepted only if steric exclusions are respected, and so on. The maximal coordination number imposed by steric exclusions is $z=12$; nevertheless, obtaining random configurations for $z>9$ is very unlikely; thus larger coordination numbers were studied on regular configurations only.

\subsection{3D Frictionless grains} 

In frictionless systems there is one variable per contact, and the constraints for mechanical equilibrium are now three,
\begin{equation}\label{eqequ4}
  \sum_{i=1}^z f^{(n)}_{i\beta} = 0\quad , \quad \beta=x,y,z \quad ,
\end{equation}
where $f^{(n)}_{i\beta}$ is the $\beta$-th component of the normal force $\vec f^{(n)}_i$. Thus, the isostatic limit would be $z_{\rm iso}=3$. Nevertheless, such a configuration is only possible for a single unstable configuration: a planar symmetrical array of three spheres. Stable configurations are reached only for coordination numbers $z\ge z_{\rm iso}=4$. The Monte Carlo procedure starts by fixing contact positions, either regular or at random. For the initial configuration, $z-3$ normal forces are chosen randomly and set to $f^{(n)}=1$, whereas  the other three are computed from Eq. (\ref{eqequ4}). If all computed forces are repulsive, the configuration is accepted, the new pressure is computed and the Metropolis step is completed as before. 

\begin{figure}[ht]
\centering
  \includegraphics[width=8cm]{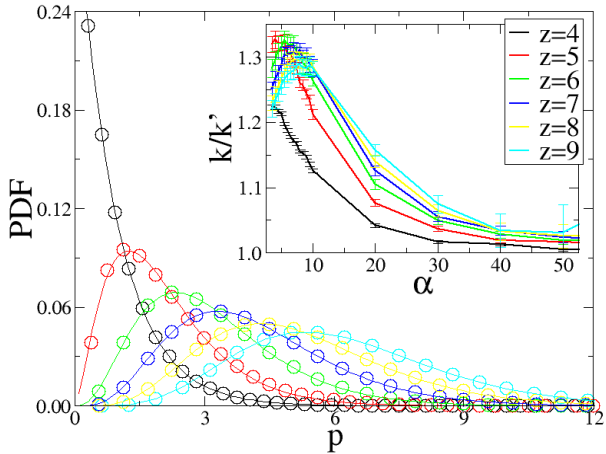}
  \caption{Probability distribution of the pressure on a single grain in a monodisperse frictionless 3D-system with  angoricity $\alpha^{-1}=0.015$ at random (circles) positions. Solid lines are Gamma functions with shape parameters $k$. (inset) Excess in shape parameter, $k-k'$ as a function of the inverse angoricity $\alpha$, with $k'=z-3$ the number of degrees of freedom in the system at random contact positions.}
  \label{fgr:PDF3D}
\end{figure}

\begin{figure}[h]
\centering
  \includegraphics[width=8cm]{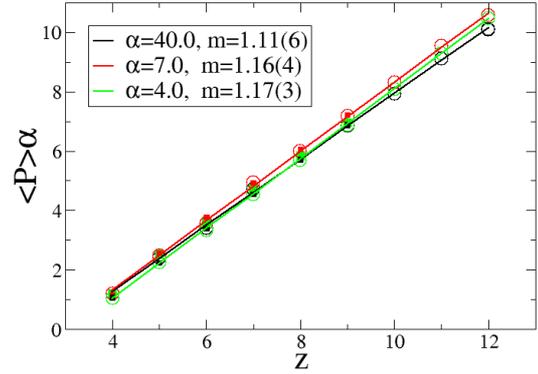}
  \caption{Average pressure $\left< p \right>$ times the inverse angoricity ($\alpha$) as a function of the coordination number $z$  for a grain in a monodisperse frictionless granular 3D-system for both fixed (empty circles) and random (filled squares) contact positions. Here, $m$ is the fitted slope.}
  \label{fgr:pa_z3}
\end{figure}

The pressure shows again gamma distributions (\textbf{Fig. \ref{fgr:PDF3D}}). At low angoricities, the parameter $k$ reaches a new value, $k = 2z - 3$, that is the number of degrees of freedom in the system. Thus Eq. (\ref{eqdis}), the probability function for a grain is now proportional to
\begin{equation} \label{eqdisp2}
  \Phi(p)\propto p^{z-3-1} \quad,
\end{equation}
and, therefore Eq. (\ref{eqpre2}), the system fulfils an equipartition relationship
\begin{equation} \label{eqest3}
   	\langle p\rangle=(z-3)\alpha^{-1} \quad.
\end{equation} 
This is exactly what is observed in our numerical simulations (\textbf{Fig. \ref{fgr:pa_z3}}).

\begin{figure}[ht]
\centering
  \includegraphics[width=8cm]{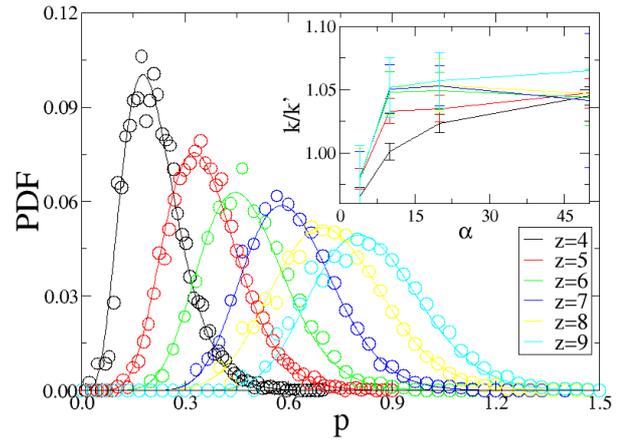}
  \caption{Probability distribution of the pressure on a single grain in a monodisperse  3D-system with static friction coefficient $\mu=0.5$ and angoricity $\alpha^{-1}=0.05$ for contacts at random positions. Solid lines are Gamma functions with shape parameters $k$.}
  \label{fgr:PDF3D_Fri}
\end{figure}

\subsection{3D Grains with static friction}

Introducing friction requires few changes. There are now three variables per contact $i$: the magnitude of the normal force $\vec f^{(n)}_i$ and the two components of the tangential force $\vec f^{(t)}_i$. The conditions for mechanical equilibrium are now six: three for the forces plus three for the torques:
\begin{equation}
  \sum_{i=1}^z f^{(n)}_{i\beta} + f^{(t)}_{i\beta}= 0\quad , \quad \beta=x,y,z  \nonumber 
\end{equation}
\begin{equation}\label{eqequ5}
	\sum_{i=1}^z \left(\vec r_i \times \vec f^{(t)}_{i}\right)_{\beta} = 0\quad , \quad \beta =x,y,z \quad ,
\end{equation}
where $\beta$ indexes the three components of each vector and $\vec r_i$ goes from the centre of the grain to the $i$-th contact point. The initial condition is chosen the same as in the 3D frictionless case (i.e. initial tangential contact forces are set to zero). 
Also, new configurations are only accepted if both all normal forces are repulsive ones and the absolute value of the tangential force at each contact does not surpass the Coulomb limit, $ f^{(t)}_i \le \mu f^{(n)}_i $.  Except by these small changes, each Monte Carlo step is completed as before.

\begin{figure}[h]
\centering
  \includegraphics[width=8cm]{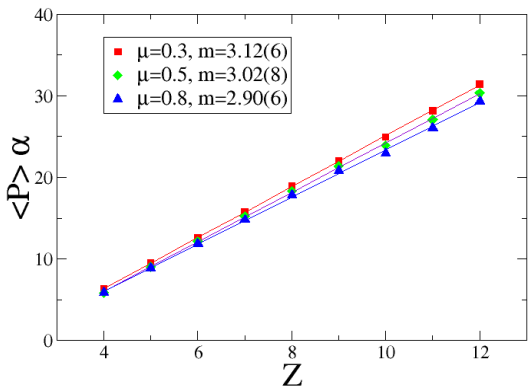}
  \caption{Average pressure $\left< p \right>$ times the inverse angoricity $\alpha$ against the coordination number $z$ for a grain in a monodisperse 3D granular system with friction. In all cases, $\alpha^{-1}=0.05$. Contact positions are chosen randomly for $z=4, 5, 6, 7, 8$ and fixed  for $z=9, 10, 11, 12$. The figure shows results for three static friction coefficients $\mu=0.3$ (squares),  $\mu=0.5$ (diamonds) and  $\mu=0.8$ (triangles)).}
  \label{fgr:pa_z4}
\end{figure}
 
The pressure histogram shows again a gamma distribution (\textbf{Fig. \ref{fgr:PDF3D_Fri}}). At low angoricities (high values of $\alpha$) the parameter $k$ goes to $k\simeq k'=3z-6$, i.e. the new number of degrees of freedom in the system. Thus, the probability function for a grain is proportional to
\begin{equation} \label{eqdisp2}
  \Phi(p)\propto p^{3z-6-1} \quad,
\end{equation}
and the system fulfills an equipartition relationship
\begin{equation} \label{eqest3}
   	\langle p\rangle=(3z-6)\alpha^{-1} \quad,
\end{equation} 
almost the same for different friction coefficients (\textbf{Fig. \ref{fgr:pa_z4}}).

\begin{figure}[h]
\centering
  \includegraphics[width=9cm]{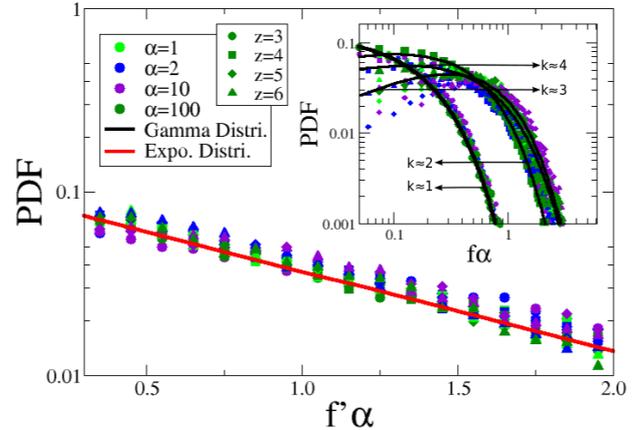}
 \caption{{\it Fundamental Forces:} Probability distribution for the  $k'=z-2$ {\it fundamental forces} $f'$ proposed for the two-dimensional frictionless case Eq. (\ref{eq:Force_Like}). Distributions are shown for several values of the inverse angoricity $\alpha$ (colors) and coordination numbers between $z=3$ and $z=6$ (shapes). The horizontal axis is scaled by $\alpha$. The solid red line corresponds to the exponential distribution Eq. (\ref{expForces}). (inset) {\it Contact forces:} Probability distribution for the original {\it contact forces} $f$, used to derive the fundamental forces. The solid black lines correspond to gamma distributions with shape parameter $k\simeq k'$, as expected for low angoricities.}
  \label{fgr:ff}
  \label{fgr:ff}
\end{figure}	

\section{Discussion and Conclusions}

We developed a Monte Carlo method to sample the ensemble of force configurations on a single grain, both with or without friction. The grain is assumed to be part of a monodisperse granular medium, either in two or  three dimensions, under isotropic compression. The contacts on the grains are chosen either at random or in a regular configuration, always respecting steric exclusions. The set of mechanical stable force configurations is sampled by a Metropolis Monte Carlo algorithm at constant angoricity. New configurations are accepted only if all normal forces are repulsive and all tangent forces are below the Coulomb's static frictional limit. 

\begin{table}[h]
\small
\centering
\caption{Product between the scale parameter $\nu$ of the gamma distribution of pressures per grain Eq. (\ref{eqksigma}) and the inverse angoricity $\alpha$ for all studied cases, $\nu \alpha^{-1}$, both in two (a) and three (b) dimensions. The magnitudes between square brackets correspond to frictional cases ($\mu=0.5$), and those without brackets, to the frictionless ones. The results support that $\nu=\alpha^{-1}$.}
\label{tableRes}

\resizebox{0.92\textwidth}{!}{\begin{minipage}{\textwidth}
\begin{tabular*}{0.4\textwidth}{@{\extracolsep{\fill}}llllll}
\cline{3-6} 
\multicolumn{2}{c}{\multirowcell{2}[0ex]{a)}} & \multicolumn{4}{c}{$\alpha$}\\ \cline{3-6} 
\multicolumn{2}{N}{}                    & \multicolumn{1}{c}{5.0} & \multicolumn{1}{c}{10.0} & \multicolumn{1}{c}{20.0} & \multicolumn{1}{c}{50.0} \\ \cline{1-6} 
\multirowcell{4}[0ex]{$Z$} 	& \multicolumn{1}{c}{3}    & 1.04 {[}1.02{]}        & 1.01 {[}1.00{]}          & 0.98 {[}0.99{]}         & 0.97 {[}0.98{]}         \\ 
				& \multicolumn{1}{c}{4}    & 1.04 {[}1.04{]}        & 1.04 {[}0.99{]}          & 0.99 {[}0.98{]}         & 0.96 {[}0.97{]}         \\ 
				& \multicolumn{1}{c}{5}    & 1.02 {[}1.04{]}        & 1.05 {[}1.00{]}          & 1.00 {[}0.99{]}         & 0.96 {[}0.96{]}         \\ 
				& \multicolumn{1}{c}{6}    & 1.04 {[}1.05{]}        & 1.05 {[}1.00{]}          & 1.00 {[}0.99{]}         & 0.95 {[}0.96{]}         \\ \cline{1-6}
\multicolumn{6}{c}{} \\ 
\multicolumn{6}{c}{} \\
\end{tabular*}
\end{minipage}}

\resizebox{0.92\textwidth}{!}{\begin{minipage}{\textwidth}
\begin{tabular*}{0.5\textwidth}{@{\extracolsep{\fill}}llllll}
\cline{3-6} 
\multicolumn{2}{c}{\multirowcell{2}[0ex]{b)}} & \multicolumn{4}{c}{$\alpha$}\\ \cline{3-6} 
\multicolumn{2}{N}{} & \multicolumn{1}{c}{5.0} & \multicolumn{1}{c}{10.0} & \multicolumn{1}{c}{20.0} & \multicolumn{1}{c}{50.0} \\ \cline{1-6} 
\multirowcell{9}[0ex]{$Z$}   & \multicolumn{1}{c}{4}          & 1.02 {[}0.92{]}                   & 0.93 {[}0.94{]}       & 0.92 {[}0.95{]}         & 0.96 {[}0.97{]}         \\ 
                             & \multicolumn{1}{c}{5}          & 1.08 {[}0.92{]}                   & 0.99 {[}0.94{]}       & 0.91 {[}0.97{]}         & 0.93 {[}0.98{]}         \\ 
                             & \multicolumn{1}{c}{6}          & 1.08 {[}0.92{]}                   & 1.02 {[}0.93{]}       & 0.93 {[}0.97{]}         & 0.93 {[}0.99{]}         \\ 
                             & \multicolumn{1}{c}{7}          & 1.09 {[}0.92{]}                   & 1.02 {[}0.93{]}       & 0.92 {[}0.96{]}         & 0.92 {[}0.99{]}         \\ 
                             & \multicolumn{1}{c}{8}          & 1.10 {[}0.93{]}                   & 1.05 {[}0.93{]}       & 0.93 {[}0.97{]}         & 0.92 {[}0.99{]}         \\ 
                             & \multicolumn{1}{c}{9}          & 1.09 {[}0.93{]}                   & 1.06 {[}0.93{]}       & 0.94 {[}0.96{]}         & 0.91 {[}0.97{]}         \\ 
                             & \multicolumn{1}{c}{10}         & 1.09 {[}0.89{]}                   & 1.07 {[}0.89{]}       & 0.95 {[}0.94{]}         & 0.93 {[}0.95{]}         \\ 
                             & \multicolumn{1}{c}{11}         & 1.09 {[}0.90{]}                   & 1.07 {[}0.90{]}       & 0.96 {[}0.94{]}         & 0.94 {[}0.95{]}         \\ 
                             & \multicolumn{1}{c}{12}         & 1.09 {[}0.91{]}                   & 1.08 {[}0.91{]}       & 0.97 {[}0.95{]}         & 1.04 {[}-{]}            \\ \cline{1-6}
\end{tabular*}
\end{minipage}}
\end{table}

Our results show, first, that the pressure on the single grain (that is, the sum of all normal forces) follows a gamma distribution $\Phi(p)$ in all cases. This coincides with the proposal of T. Aste an T. Di Matteo \cite{Aste_Emer2008} for the distribution of Vorono\"i cells in a granular media in the frame of Edward's statistical mechanics of volumes \cite{Aste_Emer2008}, which has also been confirm for grains under isotropic compression by Oquendo and co-workers \cite{Oquendo2016}. Actually, the procedure we used to find $k$ and $\nu$ are the same employed there.
The insets in \textbf{Fig. \ref{fgr:PDF1}, \ref{fgr:PDF2}, \ref{fgr:PDF3D} and \ref{fgr:PDF3D_Fri}} show that, in all cases, the scale parameter $k$ in the gamma distribution is very similar to the number of degrees of freedom $k'$ for the forces on the grain and, that $k \to k'$ in the limit for low angoricities (high values of $\alpha$). This result could be observed as a natural consequence of the sampling Metropolis Monte Carlo procedure, which moves $k'$ variables and accepts or rejects new configurations to fulfill a detailed-balance among individual configurations with exponential probability distributions. Nevertheless, the point here is that this result is not altered neither by steric exclusions nor by the restrictions on the forces (non-cohesive forces and Coulomb's static limit for the tangential forces). Moreover, the gamma distribution we postulate here for the force network ensemble is compatible with all previous descriptions \cite{Vlugt2011,Tighe2010}. 
Second, we also found in all cases (\textbf{Fig. \ref{fgr:pa_z1}, \ref{fgr:pa_zf}, \ref{fgr:pa_z3} and \ref{fgr:pa_z4}}) that the average pressure fulfills an equipartition-like relation $\langle  p \rangle = k' \alpha^{-1}$.  These two results are valid for all systems studied, either with or without friction and both in two or three dimensions. 
These findings confirm for a single grain the proposal by Tighe and Vlugt \cite{Vlugt2011} for the whole system, postulating an equipartition-like relation and that $\Phi(p) \propto p^{k'-1}$, with $k'$ the number of degrees of freedom in the system.

Actually, this two results combine to give interesting consequences. First, because in a gamma distribution the shape and scale parameters  are related by $\langle  p \rangle = k v$ Eq. (\ref{eqksigma}), it implies that scale parameter and angoricity are equal, $v=\alpha^{-1}$. Table \ref{tableRes} shows that this relation is fulfilled in all studied cases, both in 2D and 3D.
This result suggest a novel method to measure the angoricity in monodisperse granular media under isotropic compression, just by looking at the scale parameter of the distribution of pressures per grain in the ensemble Eq. (\ref{eqksigma}), resembling the propose by T. Aste and T. Di Matteo for computing the compressibility of Edward's volume ensemble for granular media from the distribution of volumes for Vorono\"i or Delaunay cells \cite{Aste_Emer2008} mentioned before.  \\
 \\
It is well known \cite{DevroyeNon1986} that the sum of $k$ independent random variables with the same exponential distribution
\begin{equation} \label{expForces}
P(f')=\alpha^{-1}\exp(-\alpha f')  
\end{equation} 
shows itself a Gamma distribution Eq. (\ref{eqdis}). That suggest that the pressure on a single grain could be considered as the sum of $k'$ independent random variables, one for each degree of freedom, with distribution Eq. (\ref{expForces}). Let us consider the 2D frictionless case, where all contact forces $f_i$ are normal. With $z$ contacts we would expect $z-2$ independent random variables with the same exponential distribution. If two contacts $a$ and $b$ are fixed at random, the $z'=z-2$ variables
\begin{eqnarray}
f'_{i}=f_i\left[1+\frac{sin(\theta_{i}-\theta_{b})-sin(\theta_{i}-\theta_{a})}{sin(\theta_{b}-\theta_{a})}\right] \quad, \\
\quad i\neq a,b \nonumber
\label{eq:Force_Like}
\end{eqnarray}
are the degrees of freedom of the system. We found that, when positive, those variables follow exactly the exponential distribution Eq. (\ref{expForces})  (\textbf{Fig. \ref{fgr:ff}}). In contrast, the $z$ contact forces $f_i$ theirselves show, for low angoricities, a Gamma distribution with shape parameter $k\simeq k'$ (\textbf{Fig. \ref{fgr:ff}}, inset), as the pressure per grain does (\textbf{Fig. \ref{fgr:PDF1}}) and, Gamma distributions grow as power laws and decay as exponentials, in agreement with previous results \cite{Radjai1990,Metzger2005}. So, the variables $f'_i$ would act as {\it fundamental forces}, similar to the fundamental volumes proposed by Aste and Di Matteo for the volume ensemble in granular media \cite{Aste_Emer2008}. If it were the general case, the force network ensemble in the limit of low angoricities would be considered as a set of $z'=N(z-z_{\rm iso})$ independent random variables with distribution Eq. (\ref{expForces}) and, all results by Tighe and Vlugt \cite{Vlugt2011} mentioned before could be derived. Moreover, by assuming that the magnitude of such elementary forces would be the only relevant variables in the system (as a first approximation for a monodisperse granular system under isotropic compression), the entropy for the force network ensemble would be
\begin{eqnarray} \label{entropy}
 S(\alpha)&=&-z'\int_{p=0}^\infty P(f')\ln\left[ P(f')/P_0 \right] df' \nonumber \\
&=&f\left[1-\ln (\alpha P_0)\right]\quad, 
\end{eqnarray}

with $p_0$ a reference pressure. If this were the general case would be an interesting topic for future research.

This work shows that the probability distribution and the equipartition relation proposed by Thige and Voigt for the pressure per grain \cite{Vlugt2011} can be obtained by extending the same assumptions to the force network ensemble on a single grain, and gives a further ground for that assumptions. Furthermore, our results suggest that, at low angoricities, not only grains can be considered independent form each other but, even more, that more elementary independent variables could exist, all with the same basic exponential distribution. These results constitute a further step in the understanding of the force network ensemble for granular media.


\bibliographystyle{spmpsci}      
\bibliography{rsc}   

%
%

\end{document}